\documentclass[conference]{IEEEtran}

\usepackage{amsmath,amsfonts}
\usepackage{algpseudocode}
\usepackage{array}
\usepackage[caption=false,font=normalsize,labelfont=sf,textfont=sf]{subfig}
\usepackage{textcomp}
\usepackage{stfloats}
\usepackage{url}
\usepackage{nameref}
\usepackage{graphicx}
\usepackage{cite}
\usepackage{physics}
\usepackage[dvipsnames,svgnames,x11names]{xcolor}
\usepackage{mathtools}
\usepackage[linesnumbered,ruled,vlined]{algorithm2e}
\usepackage{color,soul}
\usepackage[resetlabels,labeled]{multibib}
\usepackage[switch,columnwise]{lineno}
\usepackage{svg}

\usepackage[utf8]{inputenc}
\usepackage{mathtools}
\usepackage{amsmath}
\usepackage{color}
\usepackage{tikz}
\usepackage{soul}
\usepackage{changes}
\begin{document}

\title{SKI-SAT: A CMOS-compatible Hardware \\for Solving SAT Problems}

\author{
    Ahmet Yusuf Salim\textsuperscript{1}, Bart Selman\textsuperscript{2}, Henry Kautz\textsuperscript{3}, Zeljko Ignjatovic\textsuperscript{1}, and Selçuk Köse\textsuperscript{1} \\
    \textsuperscript{1}Department of Electrical and Computer Engineering, University of Rochester \\
    \textsuperscript{2}Department of Computer Science, Cornell University\\
    \textsuperscript{3}Department of Computer Science, University of Virginia at Charlottesville\\
    \texttt{asalim@ur.rochester.edu}\\
    \texttt{selman@cs.cornell.edu}\\
    \texttt{henry.kautz@virginia.edu}\\
    \texttt{zeljko.ignjatovic@rochester.edu}\\
    \texttt{selcuk.kose@rochester.edu}
}

\thanks{This work was supported in part by the Defense Advanced Research Projects Agency (DARPA) Quantum-inspired Classical Computing (QuICC) program under Air Force Research Laboratory (AFRL) contract FA8650-23-C-1034.}

\markboth{IEEE Transactions on Circuits and Systems I}
{Salim \MakeLowercase{\textit{et al.}}: SKI-SAT: A CMOS-compatible Hardware for Solving MAX-SAT Problems}

\maketitle

\begin{abstract}
Nature-inspired computation is receiving increasing attention. Various Ising machine implementations have recently been proven to be effective in solving numerous combinatorial optimization problems including maximum cut, low density parity check (LDPC) decoding, and Boolean satisfiability (SAT) problems. In this paper, a novel method is presented to solve SAT or MAX-SAT problems with a CMOS circuit implementation. The technique solves a SAT problem by mapping the SAT variables onto quantized capacitor voltages generated by an array of nodes that interact through a network of coupling units. The nodal interaction is achieved through coupling currents produced by the coupling units, which charge or discharge capacitor voltages, implementing a gradient descent along the SAT problem's cost function to minimize the number of unsatisfied clauses.
The system also incorporates a unique low-complexity perturbation scheme to avoid settling in local minima, greatly enhancing the performance of the system. 
The simulation results demonstrate that the proposed SKI-SAT is a high-performance and low-energy alternative that surpasses existing solvers by significant margins, achieving more than 10 times faster solution and 300 times less power.
\end{abstract}


\section{Introduction}

Digital computers revolutionized the world by automating complex tasks, enabling instant communication, and providing unmatched performance for data processing and artificial intelligence applications. However, recent years have seen an incredible surge in energy-intensive applications for cryptocurrency and neural network training. In digital systems information is represented and processed in one of the two discrete states while analog computers process and store information as analog values such as voltages or currents with inherently higher resolution enabling time and power-efficient solutions for certain problems. The drawback associated with analog computers is their lack of versatility, but they can outperform digital computers in certain specific functions. 
Although widespread use of analog computers for general purpose computing is unlikely, analog computers can be a favorable option for specific applications. 

In recent years, a subset of analog computation models, known as nature-inspired computation, has garnered attention for providing new avenues to solve NP-hard and NP-complete problems.
Since analytical solutions don't typically exist for these problems, digital computers are often required to rely on rather inefficient trial-and-error methods. 
An Ising machine is an emerging nature-inspired computation model that  maps a problem into an energy space. 
The system naturally creates a gradient descent to the Hamiltonian function based on its energy, allowing Ising machines to progress toward the optimal solution with the help of random perturbations.
The system is, however, likely to settle into a sub-optimal solution without a perturbation scheme\cite{zhang2022}. 
The efficiency of Ising machines in solving quadratic unconstrained binary optimization (QUBO) problems has been demonstrated to be almost a million times faster than conventional simulated annealing \cite{zhang2022} owing to the trivial mapping of QUBO to Ising formula.
However,  functions containing higher-order polynomials cannot be directly mapped to the quadratic Ising machines and therefore require preprocessing. 
The work in \cite{cilasun2024} demonstrates a methodology that first transforms a 3-SAT problem into a QUBO format, which is then decomposed into sub-problems and mapped onto a ring oscillator-based 49-spin Ising machine.
However, the auxiliary spin overhead created during the conversion of third-order terms in SAT cost function to quadratic terms in QUBO severely limits the size of 3-SAT problems that can be solved on a specific size QUBO solver (e.g., a 49-spin QUBO hardware in \cite{cilasun2024} is able to solve a 3-SAT problem with up to 20 literals and 91 clauses only by decomposing into sub-Hamiltonians). 
Therefore, supporting larger problems with more variables becomes increasingly challenging and time consuming. 
Hizzani et al. \cite{hizzani2024} report on a memristor based system that employs a Hopfield Neural Network, which eliminates the need for order reduction for third-order polynomials present in a 3-SAT problem. 
This makes the design intrinsically compatible with polynomial unconstrained binary optimization (PUBO) problems, particularly those involving SAT. 
However, the hardware implementation of this work imposes significant physical design constraints due to the quadratically growing word lines. 
This indicates that scalability for larger problems remains a challenging issue. Furthermore, Elmitwalli et al. \cite{elmitwalli2024} demonstrate how a QuBRIM system \cite{zhang2022}, originally designed to solve MaxCUT problems, can be altered to address higher-order combinatorial optimization problems, including SAT and the Traveling Salesman Problem, particularly for LDPC decoding, without the need for order reduction through multi-body interactions. 
However, the proposed system's limitation lies in the nodal interactions being constrained by the linearity of the DAC units and dynamic range for programmable resistance, which reduces the accuracy of the solution to a certain degree. 

The proposed \textbf{S}alim-\textbf{K}öse-\textbf{I}gnjatovic SAT hardware solver (SKI-SAT) offers a scheme that is not restricted by polynomial terms, as it inherently supports polynomial terms of any order. 
Additionally, the coupling precision limitations of the previous studies are no longer a concern for SKI-SAT, thanks to the binary nature of its coupling. 
Finally, SKI-SAT is easily scalable and cost-efficient thanks to the compact footprint, which is compatible with standard CMOS processes and can accommodate hundreds or even thousands of clauses.

\section{Background}\label{sec:bk}

 Solving a SAT problem can be described as answering the question of whether a boolean (i.e., logic) function evaluates true for at least one combination of the input variables. Despite the simple definition, SAT solvers have widespread applications in computer science, including electronic design automation, hardware verification \cite{vizel2015}, and cryptanalysis \cite{legendre2012}, making them a focal point of research. 
 SAT is formally an NP-complete problem, which means that it is highly likely that its general solution requires worst-case exponential time.  Modern SAT solvers use branching heuristics and dynamic programming to guide their search through the exponential space of possible solutions.  Although such techniques are often successful in practice, there remain many problems that resist solution.
Continuous advancements in algorithms and hardware have been improving SAT solver performance each year. However, the inherent computational complexity of these problems often leads to lengthy solution times, constrained by the limits of exponential-time algorithms in a search space as broad as $2^N$ combinations for an $N$ variable function. 
 In recent years, various Ising machine implementations have emerged to solve QUBO problems, such as MaxCUT problems. However, not all NP optimization problems can be effectively expressed in QUBO form; many include cubic or even higher-order terms. 
 These problems with higher-order terms require additional hardware, potentially reducing the effectiveness of an Ising machine when attempting to map such problems to its architecture.
 
 A SAT problem is typically presented in conjunctive normal form (CNF), where the CNF of a Boolean function contains a conjunction of clauses.
 When each clause consists of logical disconjunction of exactly $k$ literals, a SAT problem is typically termed as k-SAT problem. 
 For example, in a 3-SAT problem (i.e., $k=3$), the literals in each of the $N_C$ clauses may take any one of the N variables or their complements, as shown in Eq. (\ref{eq:literals}).

\begin{equation}
\begin{aligned}
\ell_{m,k} \in \{X_1, X_2, \ldots, X_N, \overline{X_1}, \overline{X_2}, \ldots, \overline{X_N}\}, \\
\quad \text{where} \quad 1 \leq m \leq N_C, \quad 1 \leq k \leq 3
\label{eq:literals}
\end{aligned}
\end{equation}
3-variable clauses are created by a disjunction of any three literals as shown in Eq. (\ref{eq:disjunction})
\begin{equation}
\mathcal{C}_i = (\ell_{i,1} \lor \ell_{i,2} \lor \ell_{i,3})
\label{eq:disjunction}
\end{equation}

The CNF form $F$ of a 3-SAT problem can then be written as shown in Eq. (\ref{eq:general_CNF})
\begin{equation}
F(X_1, X_2, \ldots, X_N) = \mathcal{C}_1 \land \mathcal{C}_2 \land \ldots \land \mathcal{C}_{N_C}
\label{eq:general_CNF}
\end{equation}

A generalization of the SAT problem described above is the Maximum Satisfiability problem (or MAX-SAT problem). 
Unlike a SAT problem which asks whether there is at least one assignment of all variables $X_i, 1 \leq i \leq N$ that renders all clauses $\mathcal{C}_j, 1 \leq j \leq N_C$ true, a MAX-SAT problem asks for an assignment of variables $X_i$ that maximizes the number of clauses that are made true with that assignment.

\section{Theoretical Analysis, System Design, and Circuit Level Implementation}\label{sec:SKI-SAT-embodiments}
\subsection{Theoretical Analysis} \label{sec:theory}
To construct the SKI-SAT solver machine, which refers to a CMOS circuit topology that minimizes the number of unsatisfied clauses in $F$ from Eq. (\ref{eq:general_CNF}), first an appropriate penalty or cost function is derived. For this purpose, a negation of the CNF form is considered, as shown in Eq. (\ref{eq:CNF-negation}), where the satisfiability of the CNF form $F$ is converted to the dissatisfiability of a logical Boolean function $\overline{F}$.
\begin{equation}
\overline{F}(X_1, X_2, \ldots, X_N) = \overline{\mathcal{C}_1} \lor \overline{\mathcal{C}_2} \lor \ldots \lor \overline{\mathcal{C}_{N_C}}
\label{eq:CNF-negation}
\end{equation}
The negated CNF form $\overline{F}$ is then converted into a penalty or cost function $H$ by replacing the logical OR and AND operations in $\overline{F}$ with addition and multiplication operations, respectively. Additionally, complemented variables $\overline{X_k}$ are replaced with $(1 - X_k)$. By way of example, Eq. (\ref{eq:conversion}) shows the conversion of $\overline{F}$ containing 3 clauses and 6 variables into the corresponding penalty function $H$.  
\begin{equation}
\begin{aligned}
    F &= (X_1 \lor \overline{X_2} \lor X_5) \land (\overline{X_3} \lor \overline{X_4} \lor X_5) \land (\overline{X_6} \lor X_4 \lor X_2) \\
    &\xrightarrow{\text{negation}} \\
    \overline{F} &= (\overline{X_1} \land X_2 \land \overline{X_5}) \lor (X_3 \land X_4 \land \overline{X_5}) \lor (X_6 \land \overline{X_4} \land \overline{X_2}) \\
    &\xrightarrow{\text{convert to cost}} \\
    H &= \overline{X_1} X_2 \overline{X_5} + X_3 X_4 \overline{X_5} + X_6 \overline{X_4} \overline{X_2} \\
    &= (1 - X_1) X_2 (1 - X_5) + X_3 X_4 (1 - X_5) \\
    &\quad+ X_6 (1 - X_4)(1 - X_2)
\end{aligned}
\label{eq:conversion}
\end{equation}

The next step in constructing the SKI-SAT circuit topology that minimizes penalty function $H$ is to derive a gradient vector of $H$ within the Hamming space $\{0,1\}^N$ spanned by variables $X_k, 1 \geq k \geq N$, as shown in Eq. (\ref{eq:gradH}).
\begin{equation}
    \nabla H = \Big[ \frac{\partial H}{\partial X_1} \frac{\partial H}{\partial X_2} \dots \frac{\partial H}{\partial X_N} \Big] ^T
    \label{eq:gradH}
\end{equation}

Following example in Eq. (\ref{eq:conversion}), the gradient of $H$ becomes:

\begin{equation}
\begin{aligned}
    \nabla H = \mqty[ -X_2 (1-X_5) \\ (1-X_1)(1-X_5) - X_6 (1-X_4) \\ X_4 (1-X_5) \\  X_3 (1-X_5) - X_6 (1-X_2) \\ -(1-X_1) X_2 - X_3 X_4 \\  (1-X_4)(1-X_2)]
\label{eq:example_gradH}
\end{aligned}
\end{equation}

Alternatively, partial derivatives of the cost function may be expressed in the following form:

\begin{equation}
\begin{aligned}
    \nabla H =\mqty[ -X_2 \overline{X_5} \\ \overline{X_1} \overline{X_5} - X_6 \overline{X_4} \\ X_4 \overline{X_5} \\  X_3 \overline{X_5} - X_6 \overline{X_2} \\ -\overline{X_1} X_2 - X_3 X_4 \\  \overline{X_4}\overline{X_2}]
\label{eq:example_gradH_1_2}
\end{aligned}
\end{equation}


Since all variables in Eq. (\ref{eq:example_gradH_1_2}) are logical (i.e., taking values 0 or 1),  the multiplicative terms can be replaced with AND logic gates as shown in the following Eq. (\ref{eq:example_gradH_2}):
\begin{equation}
\begin{aligned}
    \nabla H = \mqty[ -X_2 \land \overline{X_5} \\ \overline{X_1} \land \overline{X_5} - X_6 \land \overline{X_4} \\ X_4 \land \overline{X_5} \\  X_3 \land \overline{X_5} - X_6 \land \overline{X_2} \\ -\overline{X_1} \land X_2 - X_3 \land X_4 \\  \overline{X_4} \land \overline{X_2}]
\end{aligned}
\label{eq:example_gradH_2}
\end{equation}
Or alternatively, with NOR gates as shown in Eq. (\ref{eq:example_gradH_2_1}).
\begin{equation}
\begin{aligned}
    \nabla H = \mqty[ - \overline{\overline{X_2} \lor X_5} \\ \overline{X_1 \lor X_5} - \overline{\overline{X_6} \lor X_4} \\ \overline{\overline{X_4} \lor X_5} \\  \overline{\overline{X_3} \lor X_5} - \overline{\overline{X_6} \lor X_2} \\ -\overline{X_1 \lor \overline{X_2}} - \overline{\overline{X_3} \lor \overline{X_4}} \\  \overline{X_4 \lor X_2}] 
\end{aligned}
\label{eq:example_gradH_2_1}
\end{equation}

An N-dimensional real vector $\Vec{\mathbf{v}}$ is defined, which resides in a unit hypercube $\Omega_N$, meaning that for each element $v_i$ of $\Vec{\mathbf{v}}$, $v_i \in [0,1]$. In addition, a surjective function $f: \Omega_N \xrightarrow{} \{0,1\}^N$ is defined, such that the real vector $\Vec{\mathbf{v}}$ is mapped onto a Hamming vector $\mathbf{X}$, following the mapping rule in Eq. (\ref{eq:surjection}).

\begin{equation}
    \mathbf{X} = f(\vec{\mathbf{v}}) \quad \text{s.t.} \quad  
    X_i = \left\{
    \begin{array}{ll}
    0, & \text{if } v_i < 0.5 \\
    1, & \text{if } v_i \geq 0.5 
    \end{array} \right.,
    \quad 1 \leq i \leq N
    \label{eq:surjection}
\end{equation}

By following the construction rule $\frac{d}{dt} \Vec{\mathbf{v}} = -\alpha \nabla H$, a set of differential equations (\ref{eq:ODE_construction}) that govern the operation of SKI-SAT circuit is defined. 
Eq. (\ref{eq:ODE_construction}) describes the gradient decent capability of SKI-SAT in minimizing the penalty function $H$.

\begin{equation}
\frac{d v_i}{dt} = -\alpha \frac{\partial H}{\partial X_i}, \quad 1 \leq i \leq N
\label{eq:ODE_construction}
\end{equation}

Following the example of SAT problem with 3 clauses and 6 variables defined in Eq. (\ref{eq:conversion}), the corresponding set of differential equations is shown in Eq. (\ref{eq:example_ODE}).

\begin{equation}
\begin{array}{ll}
\frac{d v_1}{dt} = -\alpha \frac{\partial H}{\partial X_1}= + \alpha \overline{\overline{X_2} \lor X_5} \\ 
\frac{d v_2}{dt} = -\alpha \frac{\partial H}{\partial X_2}= -\alpha  
\overline{X_1 \lor X_5} + \alpha \overline{\overline{X_6} \lor X_4} \\ 
\frac{d v_3}{dt} = -\alpha \frac{\partial H}{\partial X_3}= -\alpha \overline{\overline{X_4} \lor X_5} \\
\frac{d v_4}{dt} = -\alpha \frac{\partial H}{\partial X_4}= - \alpha \overline{\overline{X_3} \lor X_5} + \alpha \overline{\overline{X_6} \lor X_2} \\ 
\frac{d v_5}{dt} = -\alpha \frac{\partial H}{\partial X_5}= +
\alpha \overline{X_1 \lor \overline{X_2}} + \alpha \overline{\overline{X_3} \lor \overline{X_4}} \\  
\frac{d v_6}{dt} = -\alpha \frac{\partial H}{\partial X_6}= - \alpha \overline{X_4 \lor X_2}
\end{array}
\label{eq:example_ODE}
\end{equation}

 The gradient descent nature of the system described by Eqs. (\ref{eq:ODE_construction}) and (\ref{eq:example_ODE}), and its ability to minimize the cost function $H$ (i.e., enforcing the number of dissatisfied clauses $\mathcal{C}_i$ in $F$ to be non-increasing over time) can be demonstrated by calculating the rate of change of $H$ as
 \begin{equation}
\frac{dH}{dt} = \sum_{i=1}^N \frac{\partial H}{\partial X_i} \frac{dX_i}{dt} = - \frac{1}{\alpha} \sum_{i=1}^N \frac{d v_i}{dt} \frac{dX_i}{dt} \leq 0
\label{eq:H_rate_of_change}
\end{equation}
 Since the rate of change of real variables $v_i$ and their corresponding logic variables $X_i$ are of the same polarity, the summation term to the right of Eq. (\ref{eq:H_rate_of_change}) is non-negative. 
 The resulting rate of change of $H$ is therefore non-positive (i.e., cost function $H$ is non-increasing over time).

The constant $\alpha$ in Eq. (\ref{eq:ODE_construction}) could be chosen to be equal to $I_{ref}/C$,
where $I_{ref}$ is the reference/constant current and $C$ is a capacitance value. Rearranging Eq. (\ref{eq:ODE_construction}) produces a set of equations that describe the current values into a set of capacitors $C_i$ as
\begin{equation}
I_{C_i}=C \frac{d v_i}{dt} = -I_{ref} \frac{\partial H}{\partial X_i}, \quad 1 \leq i \leq N
\label{eq:ODE_construction_with caps}
\end{equation}
 where voltage $v_i$ on a capacitor $C_i$ represents its state, which can range from $0V$ to $V_{dd}$ or a normalized range from 0 to 1. The corresponding logical variable $X_i$ is produced at the output of a comparator circuit which compares $v_i$ against a threshold $V_{th}$ which is typically $V_{dd}/2$ - or normalized to $0.5$, as previously described in Eq. (\ref{eq:surjection}).
As an example, a SKI-SAT circuit solving a 3-SAT problem with 6 variables in 3 clauses defined in Eq. (\ref{eq:conversion}) can be described as
\begin{equation}
\begin{array}{ll}
I_{C_1} = -I_{ref} \frac{\partial H}{\partial X_1}= + I_{ref} \overline{\overline{X_2} \lor X_5} \\ 
I_{C_2} = -I_{ref} \frac{\partial H}{\partial X_2}= -I_{ref}  
\overline{X_1 \lor X_5} + I_{ref} \overline{\overline{X_6} \lor X_4} \\ 
I_{C_3}  = -I_{ref} \frac{\partial H}{\partial X_3}= -I_{ref} \overline{\overline{X_4} \lor X_5} \\
I_{C_4} = -I_{ref} \frac{\partial H}{\partial X_4}= - I_{ref} \overline{\overline{X_3} \lor X_5} + I_{ref} \overline{\overline{X_6} \lor X_2} \\ 
I_{C_5} = -I_{ref} \frac{\partial H}{\partial X_5}= +
I_{ref} \overline{X_1 \lor \overline{X_2}} + I_{ref} \overline{\overline{X_3} \lor \overline{X_4}} \\  
I_{C_6} = -I_{ref} \frac{\partial H}{\partial X_6}= - I_{ref} \overline{X_4 \lor X_2}
\end{array}
\label{eq:example_ODE_Iref}
\end{equation}

\subsection{System Design and Circuit Level Implementation }
\begin{figure}[t]
\centering{\includegraphics[width=\linewidth]{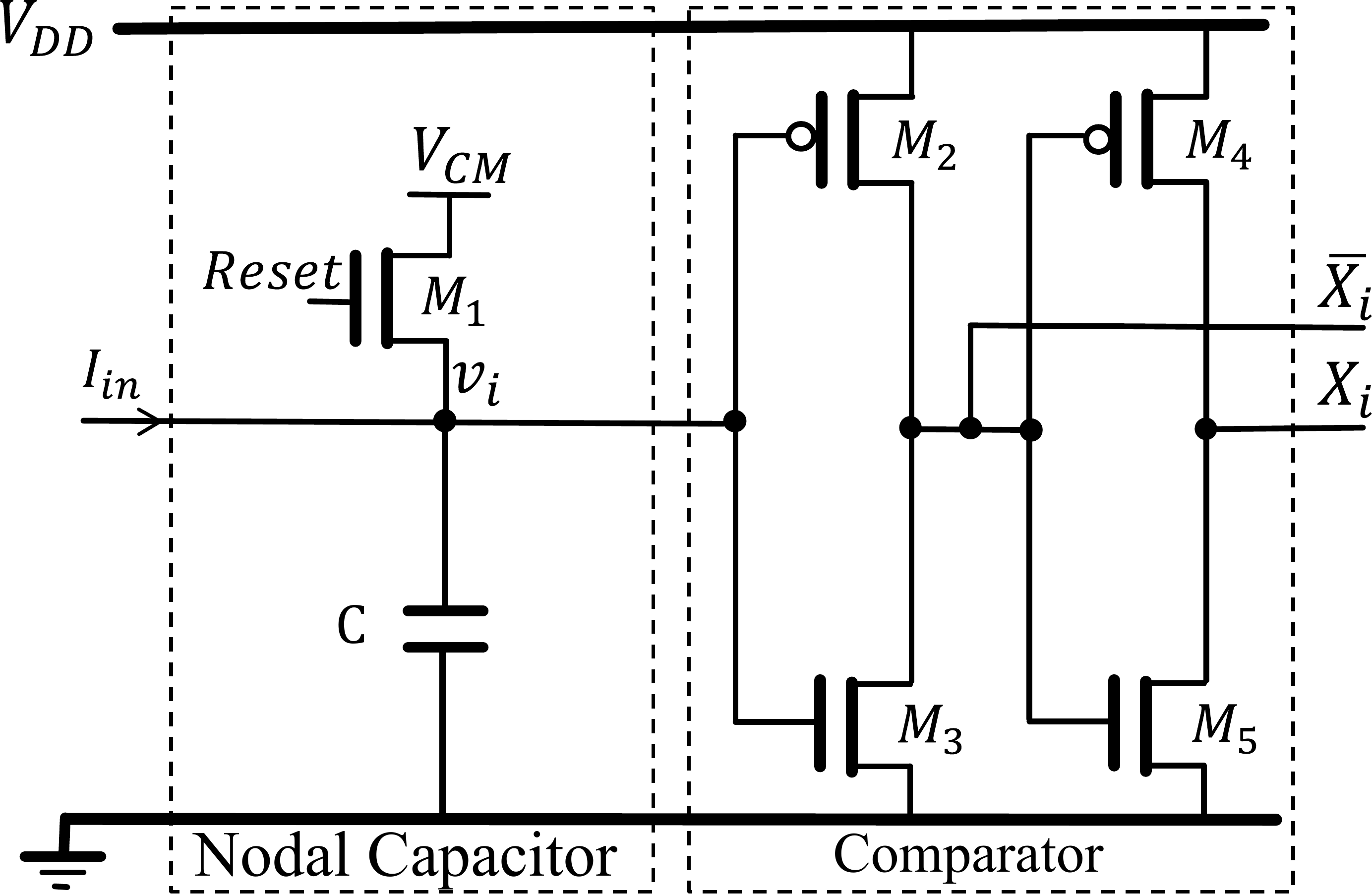}}
\caption{Example CMOS circuit for a computational node in SKI-SAT. The reset voltage $V_{CM}$ is typically chosen to be equal or close to threshold $V_{th}$ of the comparator.}
\label{fig1}
\end{figure}

\begin{figure}[t]
\centering{\includegraphics[width=\linewidth]{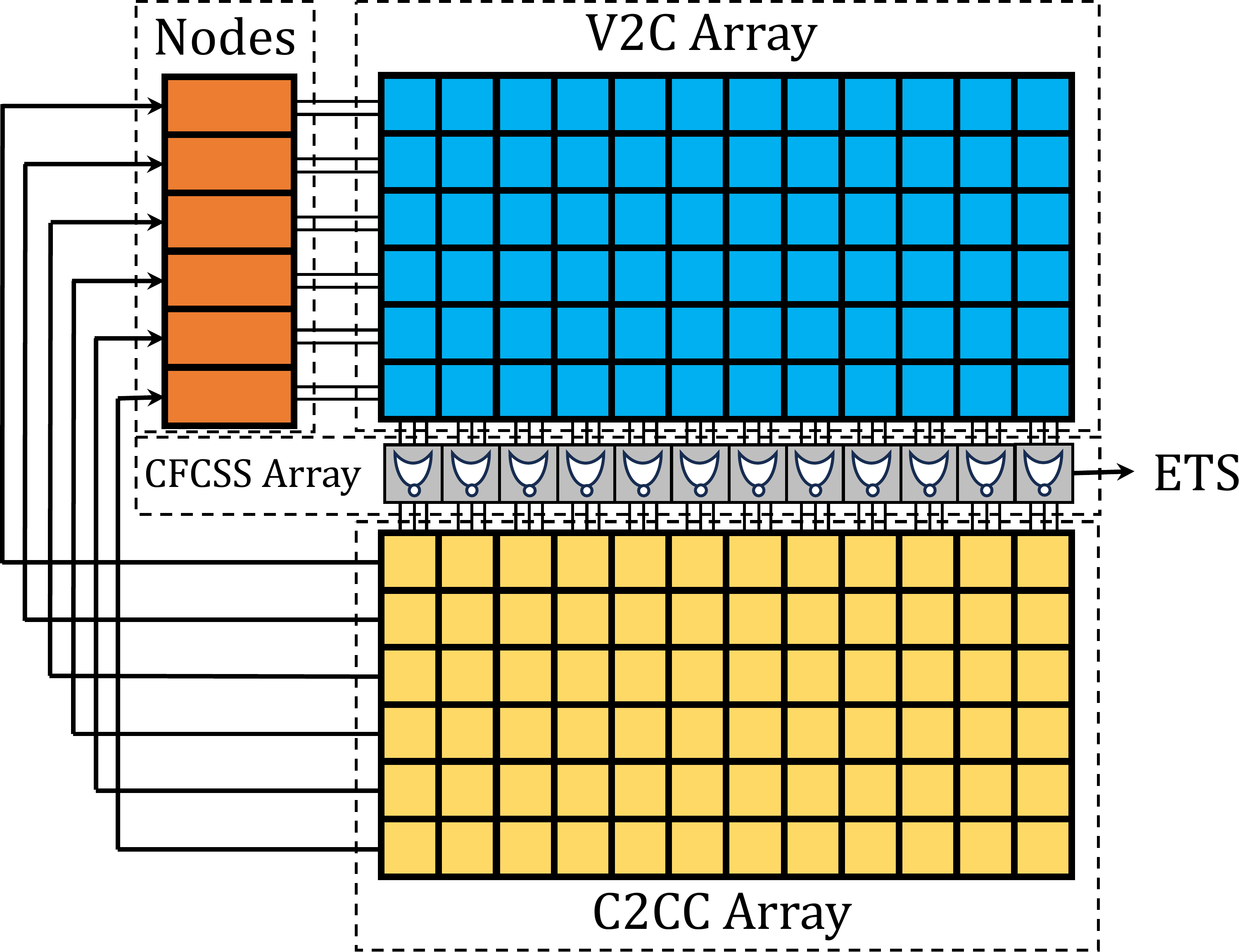}}
\caption{SKI-SAT top-level circuit architecture. }
\label{fig2}
\end{figure}

In order to construct the system governed by the described differential equations in Section \ref{sec:theory}, certain circuit topologies and their objectives are presented in this section. 
The SKI-SAT circuit topology consists of $N$ nodes for an $N$-variable function. 
Each node comprises a capacitor with voltage $v_i$ across its terminals, an input to supply charging or discharging current to the nodal capacitor, and a comparator. 
The comparator compares the capacitor voltage $v_i$ against a threshold voltage $V_{th}$, which could be set in the middle between the power rails. 
This comparison produces a binary output variable $X_i$, which can be 0V or $V_{dd}$, along with its complement $\overline{X_i}$. 
Additionally, a reset switch is placed between $V_{CM}$ and the nodal capacitor, allowing the nodal variables to be initialized to the middle rail before the system begins operation. 
An example schematic of the SKI-SAT node is shown in Fig. \ref{fig1} where the comparator is implemented as a simple inverter. 
Moreover, the two outputs ($X_i$ and $\overline{X_i}$) from this computational node  are connected to a programmable Variables-to-Clauses (V2C) array containing $N \times N_C$ units, as shown in the top-level circuit architecture of Fig. \ref{fig2}.

The V2C array provides digital outputs to clause formation and coupling control signal generation array (CFCCS) of size $1 \times N_C$ to form clauses and generate control signals associated with these clauses. 
Each unit cell in the V2C array contains memory elements to store the polarity bit as well as an address applied to the input of a digital address decoder. 
If the variable $X_i$ or its complement $\overline{X_i}$ is present in the $\mathcal{C}_j$ clause of the CNF form $F$, cell at the $(i,j)$ location in the V2C array is activated to connect the variable to the clause $\mathcal{C}_j$. 
Depending on the polarity bit stored in the memory cell of the V2C unit, either variable $X_i$ or its complement $\overline{X_i}$ is selected to be connected to the CFCCS array. 
The selection between a variable $X_i$ or its complement $\overline{X_i}$ could be achieved through a digital multiplexer (MUX) which is controlled by the polarity bit. 
The V2C unit further includes a set of output buffers that connect the variable selected through the polarity selection MUX to one of the $k$ readout lines. 
$k$ is the number of literals in the clause $\mathcal{C}_j$ that run through all rows of the $j^{th}$ column and are connected to the inputs of $j^{th}$ CFCCS unit. 
The CFCCS array is preferably located at the periphery of the V2C array as illustrated in Fig. \ref{fig2}. 

\begin{figure}[t]
\centering{\includegraphics[width=\linewidth]{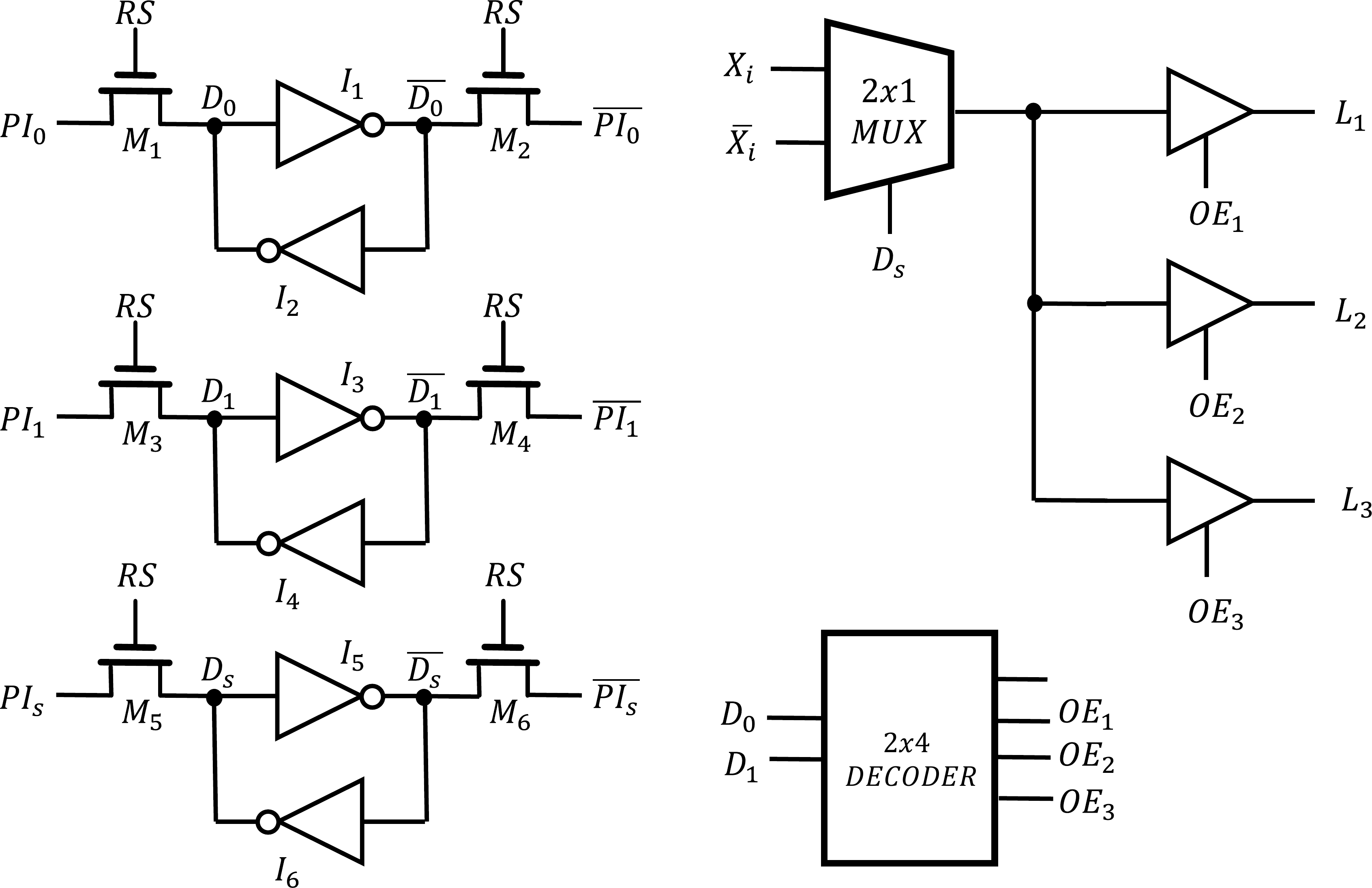}}
\caption{Example unit element of the Variables-to-Clauses $N \times N_C$ array for a 3-SAT implementation.}
\label{fig7}
\end{figure}

The output buffer in the V2C unit can be implemented with a total of $k$ 3-state inverter-based buffers with their inputs connected to the output of the polarity selection MUX and with output-enable (OE) ports connected to outputs of the address decoder. 
The digital address decoder located within the V2C cell converts the address stored in the memory of the V2C unit into at least $k$ control signals connected to the OE ports of the output buffer such that at most one of the 3-state inverter-based buffers is enabled at the time. 
Conversely, if a literal is not present in a clause, none of the OE ports are activated and the variable present at the input of the V2C cell remains disconnected from a clause. 
Fig. \ref{fig7} shows an example circuit of the V2C unit corresponding to a clause with three literals. 
The unit contains 3 memory elements e.g., latches with access switches. 
One of the memory elements is dedicated to storing the polarity selection bit $D_s$, as well as the 2-bit address bits $D_0$ and $D_1$ for readout line selection, as shown in the schematic. 
A 2-to-4 address decoder is used to either connect the variable or its complement to one of the three readout lines $L_1$, $L_2$, and $L_3$, or allow the unit and its corresponding variable to remain unconnected to any clause in the CFCCS array.

The CFCCS array contains $N_C$ units and each unit has a total of $k$ inputs. 
These inputs are connected to the $k$ readout lines of the corresponding $j^{th}$ column in the V2C array. 
Each unit in the CFCCS array contains a total of $k$ NOR logic gates, with one NOR gate for each literal in clause $\mathcal{C}_j$. 
Every NOR gate has $k-1$ inputs, which are connected to the inputs of the CFCCS unit.
A $k-1$ input NOR gate corresponding to the $m^{th}$ literal in clause $\mathcal{C}_j$ takes all literals at its inputs except the $m^{th}$ literal. 
In other words, the inputs to a $k-1$ input NOR gate corresponding to the $m^{th}$ literal are connected to all the readout lines from the $j^{th}$ column in the V2C array, except the readout line corresponding to the $m^{th}$ literal.

\begin{figure}[t]
\centering{\includegraphics[width=\linewidth]{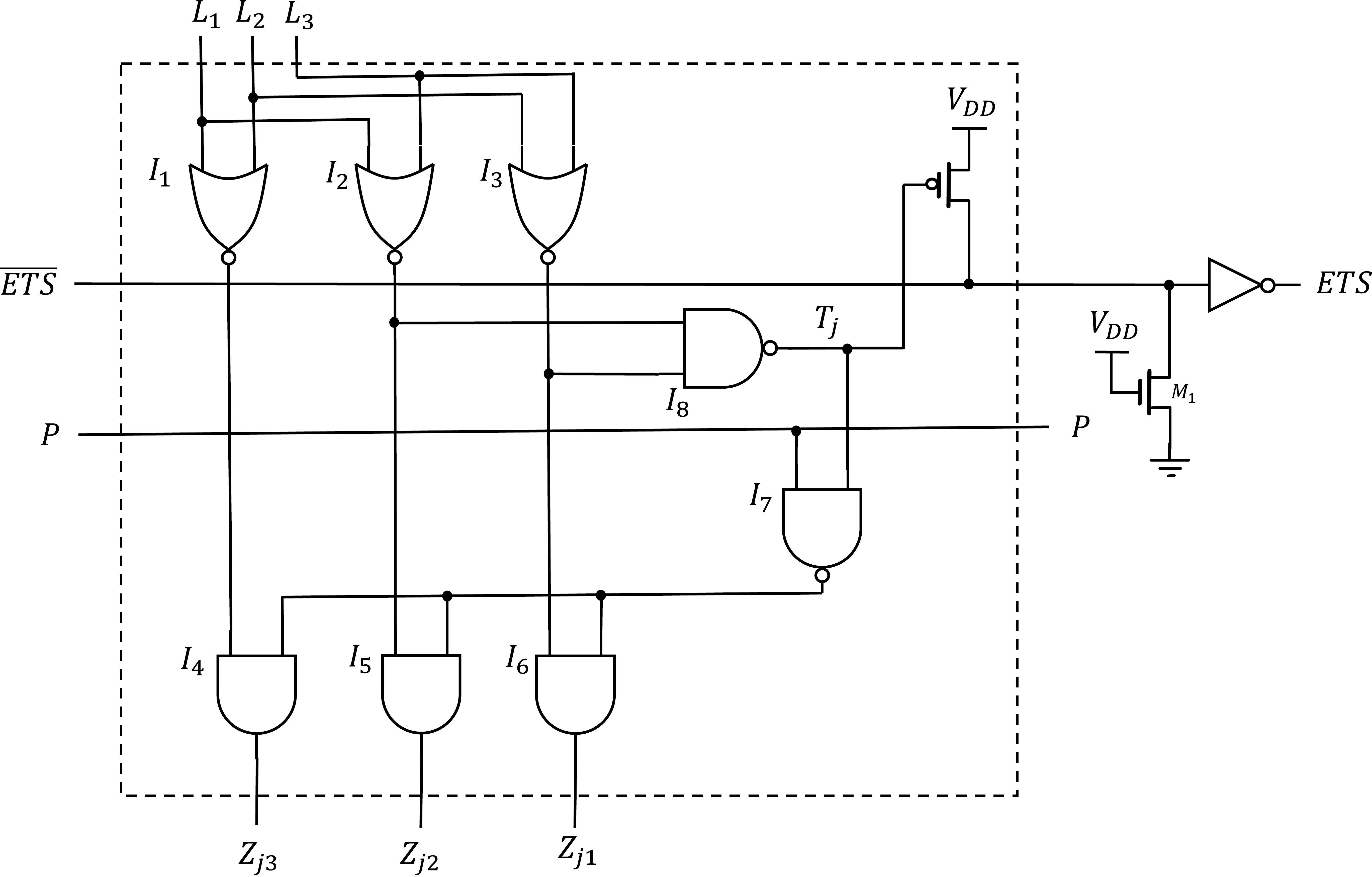}}
\caption{The CFCCS unit with clause perturbation logic and perturbation signal $P$, satisfiability signal $T_j$ and early termination signal $ETS$. }
\label{fig8}
\end{figure}

Each NOR gate in the $j^{th}$ column produces a digital output $Z_{j,m}$ (where $1 \leq m \leq k$) which is connected to Clause-to-Coupling-Current (C2CC) array of size $N \times N_c$. Each unit in the CFCCS array contains additional logic that receives outputs from the NOR gates to calculate satisfiability signal $T_j$ of clause $\mathcal{C}_j$. The logic producing satisfiability signal $T_j$ can be configured so that $T_j$ is equal to logic one if $\mathcal{C}_j$ is satisfied (or evaluates as TRUE) or $T_j=0$, otherwise. For example, in the case of a clause with 3 literals ($k=3$), the satisfiability signal $T_j$ of clause $\mathcal{C}_j$ may be formed by applying a 2-input NAND gate on the outputs from any of the two NOR gates in the $j^{th}$ CFCCS unit as shown in Fig. \ref{fig8}.

The SKI-SAT circuit described is further supplied with a perturbation circuit that perturbs the clause formation and coupling control signal generation logic in the CFCCS array. The CFCCS units are also supplied with additional logic gates (gates $I_4$ to $I_7$) that monitor the satisfiability signal $T_j$ and a perturbation control signal $P$ shared among all CFCCS units. An example CFCCS unit employing clause-perturbation logic is shown in Fig. \ref{fig8}.

To achieve perturbations in SKI-SAT, which allows for escaping local minima in the penalty function, each time a pulse in $P$ is detected by the CFCCS units, all CFCSS units whose satisfiability signals $T_j$ indicate that clause $\mathcal{C}_j$ is TRUE (or satisfied) generate logic zero on all of their output lines $Z_{j,m}$. As a result, all satisfied clauses during the pulse duration in $P$ do not contribute to the coupling currents. Therefore, these clauses do not affect the changes in nodal states $v_i$. In the absence of a pulse in $P$, all CFCCS units regardless of the value of their satisfiability signal $T_j$ contribute to the coupling currents. The pulse stream $P$ is generated so that the start of each pulse (i.e., the rising edge of the pulse) is random in time, while the duration of the pulse is fixed. In addition, the pulse stream $P$ is generated such that the average pulse rate (i.e., number of pulses per second) diminishes over the annealing time. Further details about the perturbation mechanism deployed in SKI-SAT as well as a comparison to other perturbation methods are provided in Section \ref{subsubsec:perturbation}.

The CFCCS units further incorporate logic gates to form an early termination signal ($ETS$) which serves as a trigger signal to sample the machine's current state of variables and store it as a solution found by the machine, as illustrated in Fig. \ref{fig8}. To achieve this, each CFCSS unit contains a PMOS transistor driven by the clause satisfiability signal $T_j$ which together with the PMOS transistors from all CFCCS units and the shared NMOS transistor loading the $\overline{\text{ETS}}$ readout line form an $N_C$-input NAND gate. The $\overline{\text{ETS}}$ signal is then inverted to create $ETS$.

Outputs from all AND gates (a total of $k$ outputs) in the $j^{th}$ CFCCS unit are connected to corresponding $j^{th}$ column of the Clause-to-Coupling-Current (C2CC) array containing $N \times N_c$ coupling units. For example, if clause $\mathcal{C}_j$ has three literals ($k=3)$, the $j^{th}$ CFCCS unit produces three output signals $Z_{j,m}$ where $ 1 \leq m \leq 3$ that are sent to the coupling units in the $j^{th}$ column of the C2CC array. A coupling unit at $(i,j)$ location within the C2CC array corresponds to $i^{th}$ node and $j^{th}$ clause. The coupling units are provided with an output supplying either a positive or negative reference current $I_{ref}$ or zero current depending on the polarity bit stored in the unit as well as the control signal supplied by the corresponding CFCCS units. Outputs from all coupling units in one row of the C2CC array are connected together for current summing and connected to the input of the node corresponding to that row, as illustrated in Fig. \ref{fig2}.
In addition, each coupling unit in the C2CC array contains memory elements to store polarity bit as well as an address for digital multiplexer.
For example, if the variable $X_i$ is present in the $\mathcal{C}_j$ clause as the $m^{th}$ literal, coupling unit at the $(i,j)$ location in the C2CC array is configured through its MUX to receive control signal on $Z_{j,m}$ and its polarity bit is set to logic 1. 
In this configuration, each time the control signal $Z_{j,m}$ is logic 1, the coupling unit at location $(i,j)$ provides a positive reference current $I_{ref}$ at its output. 
Otherwise, when $Z_{j,m}=0$ is received by the coupling unit $(i,j)$, its output current is set to zero. 
Likewise, in the case when a complemented variable $\overline{X_i}$ is the $m^{th}$ literal in clause $\mathcal{C}_j$, the polarity bit stored in the coupling unit $(i,j)$ is set to zero and each time $Z_{j,m}=1$ is received, the coupling unit provides a negative reference current ($-I_{ref}$) at its output. 
Otherwise, when $Z_{j,m}=0$ is received, the coupling unit produces no current at its output.
The outputs from all coupling units in the $i^{th}$ row of the C2CC array are connected together to the input of the $i^{th}$ node.

\begin{figure}[t]
\centering{\includegraphics[width=\linewidth]{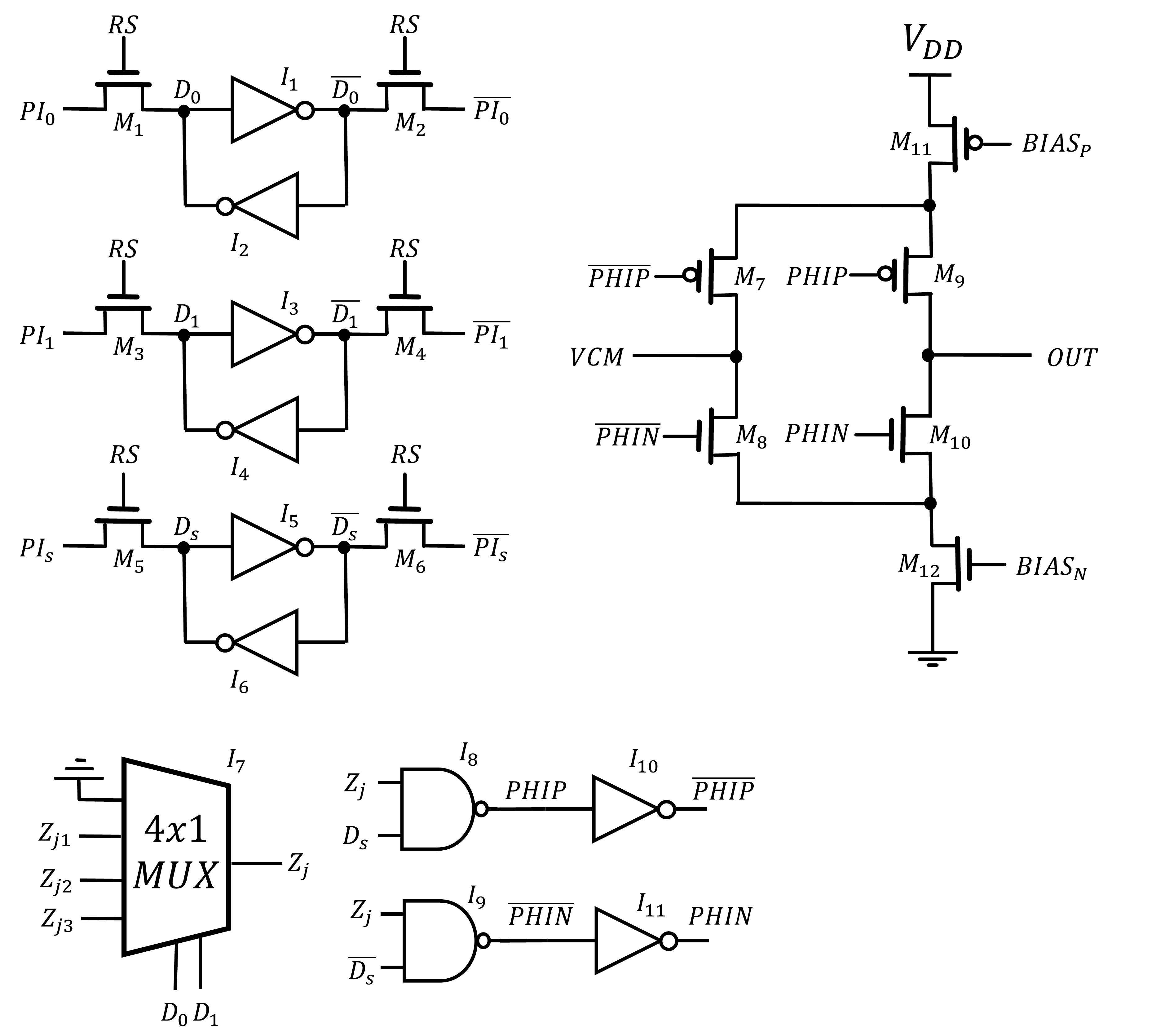}}
\caption{Clause to the coupling current (C2CC) unit for a 3-literal SAT solver implementation.}
\label{fig4}
\end{figure}
An example schematic of the C2CC unit for 3-SAT implementation is shown in Fig. \ref{fig4}. 
In this circuit, three memory elements retain information about the polarity of $D_s$ and address bits for the selection of one of the inputs  $Z_{j,m}$ supplied by the corresponding CFCCS unit in the $j^{th}$ column.
The address bits are utilized for selection through the multiplexer $I_7$. 
Depending on the polarity of $D_s$, either the NAND gate $I_8$ or $I_9$ will be activated to pull down either $PHIP$ or $\overline{PHIP}$, respectively, if $Z_j$ is set to high.
When the polarity is positive (indicating $D_s$ is high), transistors $M_9$ and $M_8$ are turned on, while $M_7$ and $M_{10}$ are turned off.
This setup connects the current source $M_{11}$ to the output, leading to a positive current at the output of the coupling unit.
Conversely, if the polarity is negative, $M_9$ and $M_8$ will turn off, and $M_7$ and $M_{10}$ will turn on, thereby connecting the current sink $M_{12}$ producing a negative current at the output. 
In the case where $Z_j$ is set to low, the switches $M_7$ and $M_8$ are engaged and connect the current source $M_{11}$ and current sink $M_{12}$ to virtual ground $V_{CM}$, while switches $M_9$ and $M_{10}$ are turned off leaving the output floating. It should be noted that switches $M_7$ and $M_8$ are not required for the operation of the coupling unit (i.e., they provide an auxiliary feature), however, employing these switches improves the linearity and settling speed of the current sources.

\section{Simulation Results and Discussion}\label{sec:rd}

This section presents the simulations performed on the proposed system and compares its performance to related work.
The performance of the proposed design is evaluated both at the circuit and behavioral levels: the former ensures the circuit's integrity, while the latter offers insights into the statistical performance of the system at larger scales. 
\subsection{Circuit-level Simulation}
\label{subsec:cir_sim}
The circuits described in the previous section are implemented in the Cadence Virtuoso environment using the TSMC 65nm process design kit.
To reduce netlist size and simulation complexity, ideal components are used for the memory units (i.e., the memory latches in V2C unit of Fig. \ref{fig7} and C2CC unit of Fig. \ref{fig4} are replaced with DC voltage sources). 
Since memory does not affect system performance and is merely required for programmability (i.e., it is accessed only once to program the SAT problem onto the hardware), using ideal components does not compromise the reliability of the simulations.
A uniform random 3-SAT Boolean function consisting of 50 literals and 300 clauses was generated to evaluate the performance of the SKI-SAT circuit solver. 
In the following step, a circuit with 50 nodes, 50 by 300 units of V2C and C2CC, and 300 units of CFCCS is instantiated and programmed using a SKILL code to implement the function. 

\begin{figure}[t]
  \centering{\includegraphics[width=\linewidth]{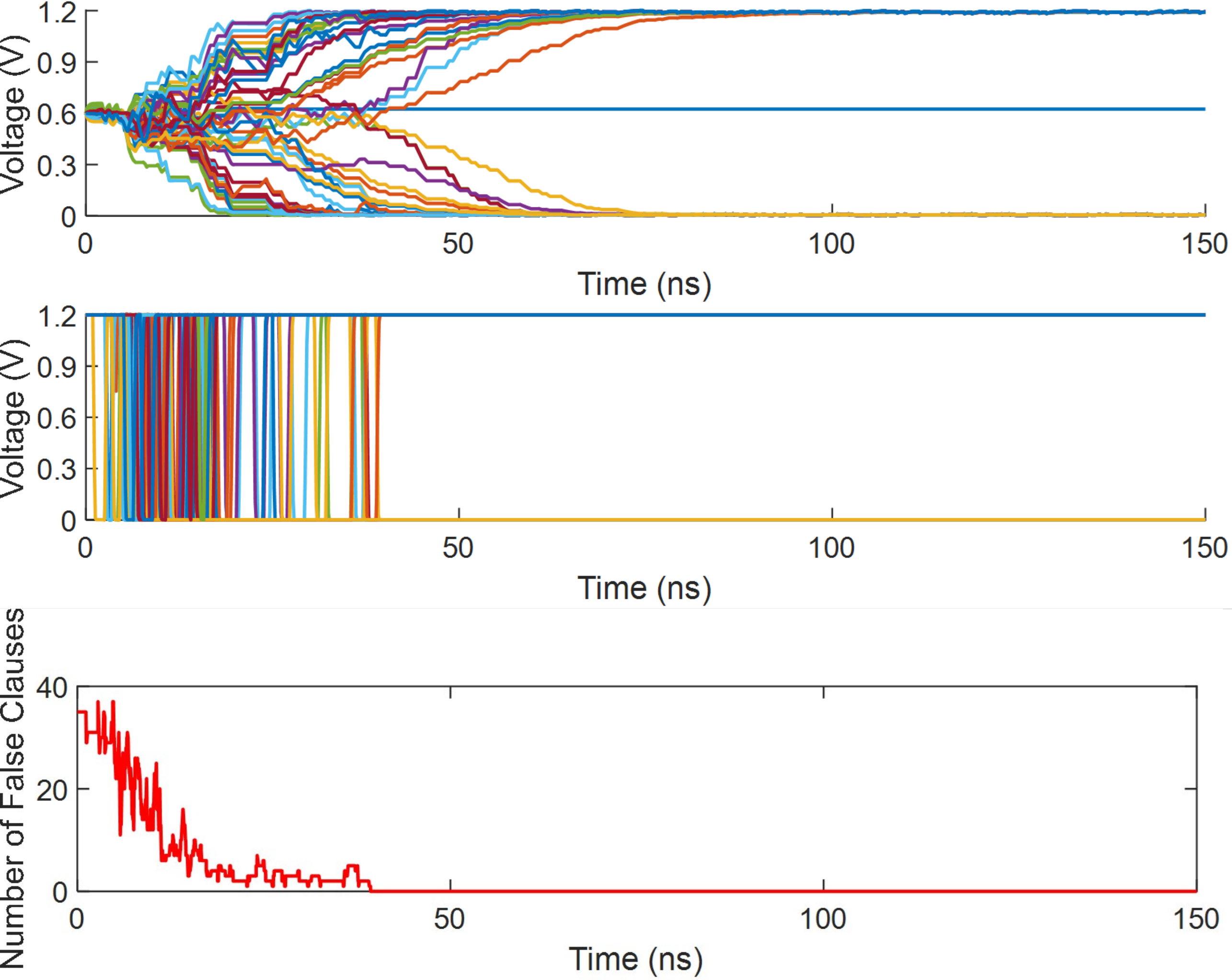}}
  \caption{A circuit-level simulation of SKI-SAT finding a solution.}
  \label{fig10}
\end{figure}

The circuit is initialized by resetting the nodal capacitors at each node to the common mode voltage, which is half the power supply — 0.6V for this process. 
The top strip in Fig. \ref{fig10} illustrates the potential difference across the nodal capacitors, starting from the middle rail and progressively diverging toward either the ground or power supply levels, with the exception of one variable, which remains in the vicinity of the middle rail while being slightly higher than the comparator's approximate threshold of $0.6V$.
This behavior is expected and commonly observed in simulations, as some variables may cease charging or discharging once the system reaches a stable equilibrium.
The middle strip of Fig. \ref{fig10} depicts the quantized outputs of the nodal capacitors, which correspond to the digital values of the literals. 
Finally, the bottom strip shows the number of false clauses gradually decreasing, eventually reaching zero, indicating that the function is satisfied with the final combination of `1's and `0's present at the nodal outputs. 
Although SKI-SAT is designed to decrease the number of unsatisfied clauses over time through its gradient decent nature, perturbations are needed to escape local minima in search for a variable assignment that maximizes the number of satisfied clauses. 
The particular perturbation scheme used in SKI-SAT introduces a ``don't care" state, during which the currents from satisfied clauses are disconnected from nodal capacitors.
Temporarily disconnecting the satisfied clauses from contributing to the charging currents effectively allows the system to enforce an assignment of variables that would satisfy the remaining unsatisfied clauses with disregard to the already satisfied clauses, which might push the system to a higher energy state.
This behavior is critical to maintain a robust solver that avoids getting stuck at local minima, which is a common issue for all stochastic solvers. 
For this particular function, the system reaches the global minimum within 40 $ns$, consuming an average current of 31.4 $mA$ during the annealing period from a 1.2 V supply, resulting in 37.68 $mW$ power consumption, excluding the memory elements and perturbation unit.

\begin{figure}[t]
  \centering{\includegraphics[width=\linewidth]{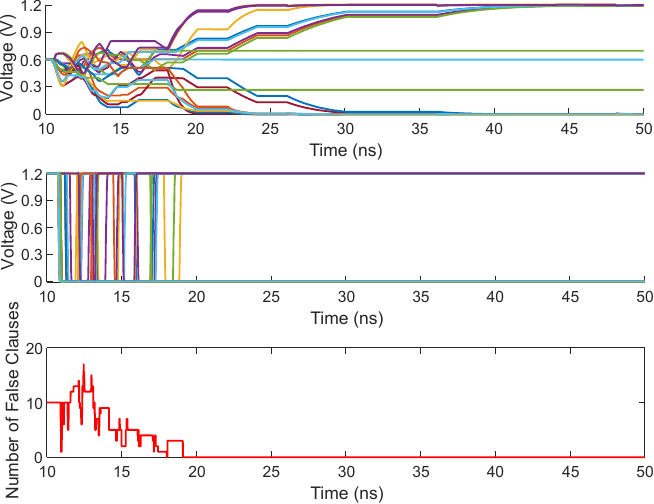}}
  \caption{SATLIB instance uf20-91/014 settling into a correct assignment.}
  \label{fig12}
\end{figure}

SATLIB \cite{SATLIB}, a widely adopted benchmark, is utilized in this paper due to its incorporation of hard randomly generated formulas; such formulas arise when the ratio of clauses to variables is at a certain  critical ratio \cite{selman1996}. 
A similar experiment was run on the SATLIB \cite{SATLIB} instance uf20-91/014 for the purposes of demonstrating the effectiveness of SKI-SAT on a commonly used benchmark item. 
Fig. \ref{fig12} shows the 20 variables diverging and eventually settling into one of the global minima, thereby finding a satisfying assignment. 
For brevity, the initialization phase during the first 10 ns is omitted.

\begin{figure}[t]
  \centering{\includegraphics[width=\linewidth]{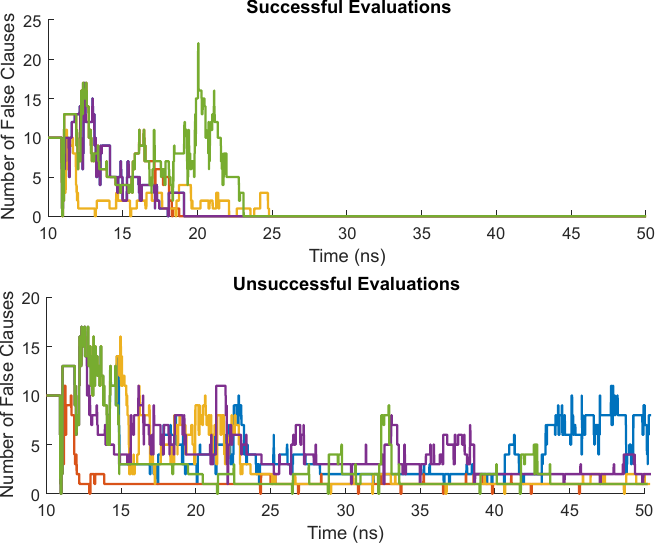}}
  \caption{Penalty function over time for a 20-variable 3-SAT instance (uf20-91/014) as solved by SKI-SAT solver for 10 iterations resulting in $50\%$ success rate. \textit{Top trace}: 5 iterations resulting in a successful evaluation - i.e., global minimum corresponding to zero unsatisfied clauses is found. \textit{Bottom trace}: 5 unsuccessful iterations where the machine remains in the vicinity of local minima.}
  \label{fig13}
\end{figure}

Stochastic solvers inherently possess a probabilistic chance of solving a given function, and SKI-SAT is not an exception. 
To analyze the statistical properties of such solvers, it is necessary to perform a large number of runs. 
Circuit simulations, however, are notoriously computationally expensive and call for realistic simplifications. 
With this in mind, circuit simulations are only carried out on SAT problems with a smaller number of variables (e.g., up to 20) and a smaller number of runs. 
The obtained statistical results are used to verify the validity of the SKI-SAT behavioral model developed in MATLAB, which is then used to estimate SKI-SAT performance for large scale SAT problems.
Fig. \ref{fig13} demonstrates five successful evaluations and five unsuccessful evaluations out of ten iterations with different perturbation sequences for the uf20-91/014, resulting in a success rate of 0.5 for SKI-SAT for this particular function.

 Furthermore, simulations show that the absence of perturbations causes the solver to become stuck in a local minimum due to the greedy gradient descent even when the transient noise is included in simulations. 
 This observation highlights the importance of random perturbations to system performance. 
 In addition, including the transient noise in simulations do not change the success rate, demonstrating the proposed circuit's resilience against inherently present electronic noise.
 Further, certain instances from SATLIB are quite trivial such that the correct solution is found rapidly without any perturbation. 
 However, such instances are rare and unlikely to exist for instances with larger energy landscapes that contain far more literals and clauses. 

\subsection{High-level Model}
\label{subsec:high_sim}
The SKI-SAT behavioral model is implemented in MATLAB as a fixed-step solver with discrete states. 
The fixed time step of $\Delta t = 20 ps$ is chosen for this experiment. 
In each time step, discrete voltage values on nodal capacitors are incremented or decremented by an integer multiple of $\Delta V$. 
Assuming the nodal capacitance value of 200 fF and a reference current of 10 $\mu$A (i.e., current produced by the current source $M_{11}$ and current sink $M_{12}$ of the C2CC unit in Fig. \ref{fig4}), the voltage increment $\Delta V$ is determined as

\begin{equation}
\Delta V = \frac{I_{\text{ref}} \times \Delta t}{C} = 1 mV
\label{eq:Delta}
\end{equation}

\begin{figure}[t]
  \centering{\includegraphics[width=0.8\linewidth]{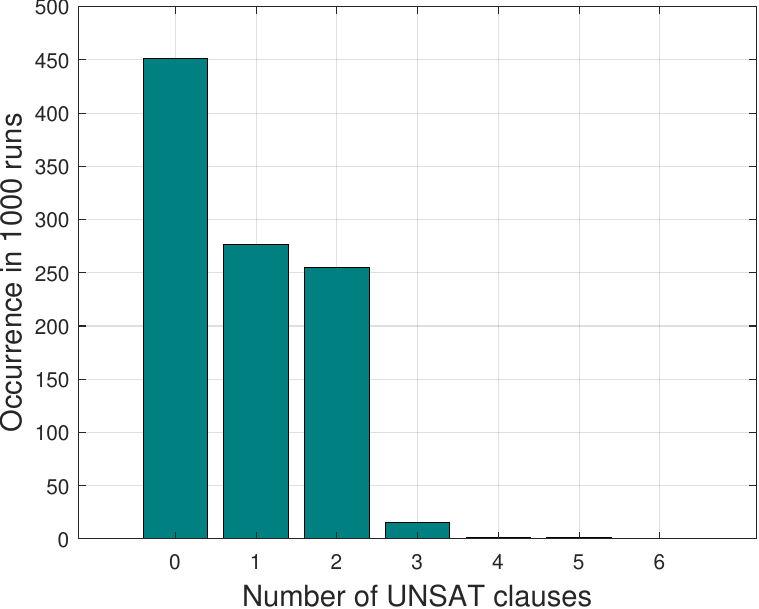}}
  \caption{Solution histogram for 1,000 iterations of SKI-SAT MATLAB model solving uf20-91/014 with $45.1\%$ success rate.}
  \label{fig14}
\end{figure}

The MATLAB script starts with fetching the Boolean function and initializes two sets of variables for the analog voltage values across the nodal capacitors and digitized outputs of the literals. 
In this model, the capacitor voltages are initialized to mid-rail value of $V_{dd}/2$ with additional random component to account for $kT/C$ noise present in the circuit. 
Additionally, a random perturbation sequence is generated, clocked at a period of 320 ps, meaning the perturbation may change state every 16 simulation steps, establishing a realistic clocking speed. 
The code keeps track of how many clauses remain unsatisfied at each step and records the first step where all clauses are satisfied while simulating a system where the capacitor voltages are updated based on the satisfaction of clauses at every step. 
The model utilizes discrete voltage increments and decrements to simulate analog behavior. 
The literal values are updated by comparing the voltage values against the threshold. 
The system is in greedy mode when there is no perturbation and tries to diminish the number of unsatisfied clauses. 
On the other hand, the perturbation stops the greedy behavior and lets some of the satisfied clauses break by aborting the inputs from satisfied clauses. 
The perturbation increases the likelihood of the system eventually descending into global minima. 
Running the uf20-91/014 for 1,000 repeats with the described behavioral model reveals the success rate of $45.1 \%$, as shown in Fig. \ref{fig14}, which is fairly consistent with the $50\%$ success rate obtained through circuit simulations.

\subsection{Perturbations in SKI-SAT}
\label{subsubsec:perturbation}
A key feature of the SKI-SAT is related to how it escapes local minima through the utilization of random perturbations. 
The primary goal of perturbation is to allow the system to occasionally choose a random path through the energy landscape, which does not align with the steepest gradient descent direction and does not necessarily lead to a reduction in the number of unsatisfied clauses. 
The SKI-SAT employs an highly-effective and resource-efficient solution to achieve this through the use of a single and global random (or pseudo-random) signal $P$. 
As shown in Fig. \ref{fig8}, the CFCCS units generate coupling signals $Z_{j,m}$ either to change variable assignments to achieve satisfiability or, if only one variable satisfies the clause, to maintain its current state.
When a pulse is detected at $P$, the coupling signals $Z_{j,m}$ for satisfied clauses cease, allowing only unsatisfied clauses to generate coupling signals.
During this phase, some of the satisfied clauses are expected to become unsatisfied while previously unsatisfied clauses will turn into satisfied, which will ultimately assist the system in finding the global minimum. 
The truth table for the CFCCS unit that summarizes the function of the perturbation signal $P$ is depicted in Table \ref{tab:truth_table}.
The perturbation signal $P$ is configured such that the perturbation pulse length is constant (e.g., $320ps$) and the pulse density (i.e., probability of entering perturbation mode) decays linearly over time.
\begin{table}[!t]
    \caption{Truth table of CFCCS unit showing inputs $P$, $L_1$, $L_2$, $L_3$, and corresponding outputs $T_j$, $Z_{j1}$, $Z_{j2}$, and $Z_{j3}$.}
    \begin{center}
        \scriptsize 
        \begin{tabular}{|c|c|c|c!{\vrule width 1.5pt}c|c|c|c|}
            \hline
            \multicolumn{4}{|c!{\vrule width 1.5pt}}{\textbf{Inputs}} & \multicolumn{4}{c|}{\textbf{Outputs}} \\
            \hline
            $P$ & $L_1$ & $L_2$ & $L_3$ & $T_j$ & $Z_{j1}$ & $Z_{j2}$ & $Z_{j3}$ \\
            \hline
            0 & 0 & 0 & 0 & 0 & 1 & 1 & 1 \\
            0 & 0 & 0 & 1 & 1 & 0 & 0 & 1 \\
            0 & 0 & 1 & 0 & 1 & 0 & 1 & 0 \\
            0 & 0 & 1 & 1 & 1 & 0 & 0 & 0 \\
            0 & 1 & 0 & 0 & 1 & 1 & 0 & 0 \\
            0 & 1 & 0 & 1 & 1 & 0 & 0 & 0 \\
            0 & 1 & 1 & 0 & 1 & 0 & 0 & 0 \\
            0 & 1 & 1 & 1 & 1 & 0 & 0 & 0 \\
            \hline
            1 & 0 & 0 & 0 & 0 & 1 & 1 & 1 \\
            1 & 0 & 0 & 1 & 1 & 0 & 0 & 0 \\
            1 & 0 & 1 & 0 & 1 & 0 & 0 & 0 \\
            1 & 0 & 1 & 1 & 1 & 0 & 0 & 0 \\
            1 & 1 & 0 & 0 & 1 & 0 & 0 & 0 \\
            1 & 1 & 0 & 1 & 1 & 0 & 0 & 0 \\
            1 & 1 & 1 & 0 & 1 & 0 & 0 & 0 \\
            1 & 1 & 1 & 1 & 1 & 0 & 0 & 0 \\
            \hline
        \end{tabular}
    \label{tab:truth_table}
    \end{center}
\end{table}

\begin{figure}[t]
  \centering{\includegraphics[width=0.8\linewidth]{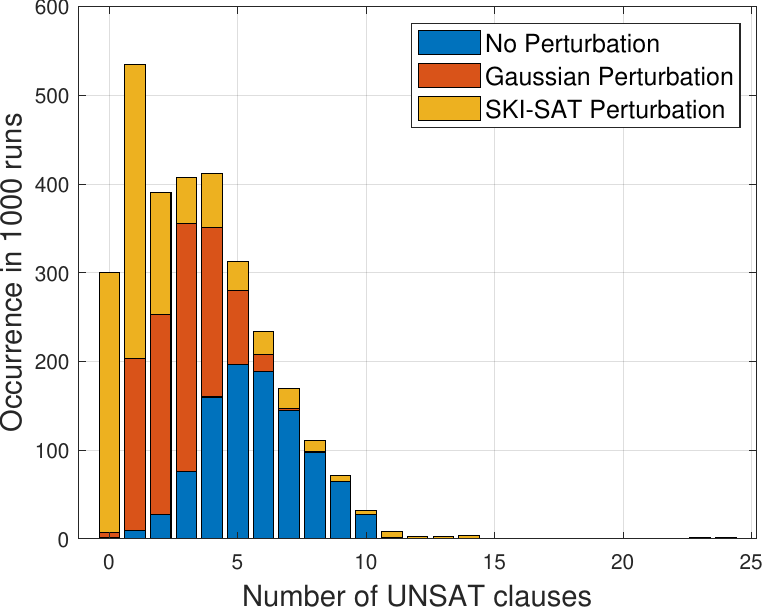}}
  \caption{SKI-SAT perturbation is effective in improving the success rate.}
  \label{fig11}
\end{figure}

In order to demonstrate the significance of the perturbations to system performance, various perturbation scenarios have been considered. 
A SATLIB instance, specifically uf50-218/0100, is simulated using the MATLAB model 1,000 times with three different approaches. 
First, when no perturbations are introduced forcing the machine to rely solely on its gradient descent in search for a global minimum, the success rate was quite low, with only 2 successful runs out of 1,000. 
Second, an alternative perturbation strategy is implemented, where $8mV_{RMS}$ Gaussian-distributed noise is randomly injected into the nodal capacitors. 
The RMS value of $8mV_{RMS}$ for injected noise is determined as optimal to maximize the success rate in this experiment. 
This approach improves the outcome, yielding 5 successful cases with zero unsatisfied clauses at the end of the annealing periods. 
Finally, with SKI-SAT's own perturbation mechanism, a success rate of 29.3$\%$ is achieved. 
The efficiency of the SKI-SAT perturbation as compared to the alternative method and the no-perturbation case is illustrated in Fig. \ref{fig11}.

\subsection{Performance Comparison}
\label{subsubsec:comparison}
A number of selected benchmark instances from SATLIB are used for comparison between a hardware implementation of AmoebaSAT\cite{hara2019}, WalkSAT\cite{selman1994}, and SKI-SAT. 
AmoebaSAT \cite{aono2013} is a bio-inspired algorithm that has been demonstrated to solve SAT problems using nanoelectromechanical devices. 
Hara et al. \cite{hara2019} report the clock frequency and average number of cycles required for solution over 100 repeats of an FPGA implementation of AmoebaSAT. 
The provided information allows the calculation of the total time to solution (TtS) by multiplying the clock period by the average number of cycles. 
Furthermore, the latest version of the WalkSAT solver, found in the GitLab repository \cite{kautzWalksat}, is executed with the default settings on an Apple Silicon M1 CPU with 8 GB of RAM. 
The WalkSAT algorithm sets a predefined number of steps, known as the cutoff number, after which it either finds a solution or exits the iteration and restarts with a new initial assignment. 
Sweeping the cutoff parameter shows that the default cutoff setting provides a good balance. 
The instances tested for this paper neither suffer from a low cutoff rate, which leads to a low success rate, nor from an excessively large limit that causes the solver to get stuck in a cycle, taking millions of flips to escape without success. 
However, it is not guaranteed that WalkSAT will find a solution in every trial. 
Therefore, it is necessary to calculate the TtS in a way that accounts for unsuccessful cases. 
The number of repetitions required to solve a function with a $99\%$ probability given a certain success rate ($SR$) is determined as
\begin{equation}
N_{99\%} = \left\lceil \frac{\log_{10}(0.01)}{\log_{10} \left( 1 - SR \right)} \right\rceil
\label{eq:TTS_99}
\end{equation}
The average time per assignment multiplied by the number of runs needed to solve an instance with $99\%$ probability is used to estimate TtS for WalkSAT as shown in\cite{Hamerly2019}.
Lastly, the SKI-SAT behavioral model in MATLAB is employed to solve the same instances. 
This model is used to project solution times for SKI-SAT, taking into account the success rate in the same manner as for WalkSAT. 
Table \ref{tab:benchmark_comparison} summarizes the benchmarking results for the three methods. 
\begin{table}[!t]
    \caption{Comparison of TtS for HW AmoebaSAT, WalkSAT, and SKI-SAT results for different benchmarks in units of $\mu s$.}
    \begin{center}
        \scriptsize 
        \begin{tabular}{|c|c|c|c|}
            \hline
            \textbf{Benchmark}  & \textbf{HW AmoebaSAT \cite{hara2019}} & \textbf{WalkSAT \cite{kautzWalksat}} & \textbf{SKI-SAT} \\
            \hline
            uf50-218/0100   & 4.13  & 187  & 4.2 \\ 
            \hline
            uf50-218/0410   & 4.36  & 141  & 4.2 \\ 
            \hline
            uf50-218/0767   & 8.30  & 365  & 10.8 \\ 
            \hline
            uf100-430/0285  & 356    & 5872  & 344.7 \\ 
            \hline
            uf150-645/0100  & 1832   & 6026  & 137.7 \\ 
            \hline
            uf225-960/028   & 3078   & 1508  & 27.6 \\ 
            \hline
        \end{tabular}
    \label{tab:benchmark_comparison}
    \end{center}
\end{table}

The SKI-SAT outperforms the advanced WalkSAT, achieving solution times that are more than 10 times faster.
In addition, a crucial figure of merit to compare different solvers is the energy to solution (EtS). 
EtS is determined by multiplying the time to solution (TtS) with the power consumption (P) to determine the energy required to solve a function as
\begin{equation}
EtS = TtS \times P
\label{eq:ETS}
\end{equation}

While WalkSAT solvers executed on von Neumann computing platforms consume energy on the order of Watts, the SKI-SAT circuit described in Section \ref{subsec:cir_sim} consumes about 20 $mW$(including perturbation sequence generation logic) while solving the uf20-91/014 3-SAT instance. 
Apple's M1 CPU consumes about 7.5 $W$ of power while executing the WalkSAT solver for the same instance. In order to estimate the power consumption, the powermetrics feature~\cite{powermetrics} is used to sample the average power drawn by the CPU with certain intervals. The idle CPU power is then subtracted from the CPU power observed while WalkSAT was running. It should be noted the EtS is slightly better when the CPU is operated in low power mode, however, this case yields a curtailed performance in terms of TtS. 
Consequently, the EtS figure of merit for SKI-SAT is up to thousands of times better than that of WalkSAT, offering a high-performance and low-power hardware alternative.

\section{Conclusions} \label{sec:c}
A new power- and area-efficient SAT solver, SKI-SAT, is proposed in this paper that is capable of solving problems with more than quadratic terms and optimized for seamless CMOS implementation.
The work demonstrates a highly scalable architecture that inherently supports third-order polynomials found in SAT problem's cost functions. The circuit implementation validates the architecture as an effective solution for SAT problems, requiring over 300 times less power. 
The behavioral model shows that SKI-SAT achieves, on average, a 38-fold reduction in solution time for the selected benchmark instances. Finally, the key performance metric, energy to solution (EtS), is orders of magnitude better than that of conventional SAT-solving algorithms.
In conclusion, the proposed SKI-SAT solver demonstrates significant improvements in power and area efficiency, effectively addressing problems with more than quadratic terms while supporting seamless CMOS implementation.

\section{Acknowledgments} \label{sec:ack}
This work was supported in part by the Defense Advanced Research Projects Agency (DARPA) Quantum-inspired Classical Computing (QuICC) program under Air Force Research Laboratory (AFRL) contract FA8650-23-C-1034.
\bibliographystyle{IEEEtran}
\bibliography{Main_rev1}

\end{document}